\documentclass[referee]{aa} 

\bibpunct{(}{)}{;}{a}{}{,}

\usepackage{graphicx}
\usepackage[varg]{txfonts}
\begin{document}

\title{A bright-rimmed cloud sculpted by the \ion{H}{ii} region Sh2-48} 
\subtitle{}

\author {M. E. Ortega \inst{1}
\and S. Paron \inst{1,2,3}
\and E. Giacani \inst{1,2}
\and M. Rubio \inst{4}
\and G. Dubner \inst{1}
}

\institute{Instituto de Astronom\'{\i}a y F\'{\i}sica del Espacio (IAFE),
             CC 67, Suc. 28, 1428 Buenos Aires, Argentina\\
             \email{mortega@iafe.uba.ar}
\and FADU - Universidad de Buenos Aires, Ciudad Universitaria, Buenos Aires
\and CBC - Universidad de Buenos Aires, Ciudad Universitaria, Buenos Aires
\and Departamento de Astronom\'{\i}a, Universidad de Chile, Casilla 36-D, Santiago, Chile
}

\offprints{M. E. Ortega}

   \date{Received <date>; Accepted <date>}

\abstract {}{To characterize a bright-rimmed cloud embedded in the
  \ion{H}{ii} region Sh2-48 searching for evidence of triggered star
  formation.}{We carried out observations towards a region of
  2$^{\prime} \times 2^{\prime}$~centered at RA=18$^{\mathrm h}$
  22$^{\mathrm m}$ 11.39$^{\mathrm s}$ , dec.=-14$^{\circ}$
  35$^{\prime}$ 24.81$^{\prime\prime}$(J2000) using the Atacama
  Submillimeter Telescope Experiment (ASTE; Chile) in the $^{12}$CO
  J=3--2, $^{13}$CO J=3--2, HCO$^+$ J=4--3, and CS J=7--6 lines with
  an angular resolution of about 22$^{\prime\prime}$. We also present
  radio continuum observations at 5~GHz carried out with the Jansky
  Very Large Array (JVLA; EEUU) interferometer with a synthetized beam
  of 7$^{\prime\prime}\times 5^{\prime\prime}$. The molecular
  transitions are used to study the distribution and kinematics of the
  molecular gas of the bright-rimmed cloud. The radio continuum data
  is used to characterize the ionized gas located at the illuminated
  border of this molecular condensation. Combining these observations
  with infrared public data allows us to build up a comprehensive
  picture of the current state of star formation within this
  cloud.}{The analysis of our molecular observations reveals the
  presence of a relatively dense clump with $n({\rm H}_2) \sim 3
  \times 10^3 {\rm cm}^{-3}$, located in projection onto the interior
  of the \ion{H}{ii} region Sh2-48. The emission distribution of the four
  observed molecular transitions has, at $V_{LSR} \sim$ 38~kms$^{-1}$,
  morphological anti-correlation with the bright-rimmed cloud as seen
  in the optical emission. From the new radio continuum observations
  we identify a thin layer of ionized gas located at the border of the
  clump which is facing to the ionizing star. The ionized gas has an
  electron density of about 73 cm$^{-3}$ which is a factor three
  higher than the typical critical density ($n_c \sim 25~{\rm
    cm}^{-3}$) above which an ionized boundary layer can be formed and
  be maintained. This fact supports the hypothesis that the clump is
  being photoionized by the nearby O9.5V star, BD-14 5014. From the
  evaluation of the pressure balance between the ionized and molecular
  gas, we conclude that the clump would be in a pre-pressure balance
  state with the shocks being driven into the surface layer. Among the
  five YSO candidates found in the region, two of them (class I), are
  placed slightly beyond the bright rim suggesting that their
  formation could have been triggered via the radiation-driven
  implosion process.}{}

\keywords{Stars: formation -- ISM: clouds -- (ISM): \ion{H}{ii} regions}

\maketitle 

\begin{figure*}
\centering
\includegraphics[width=17cm]{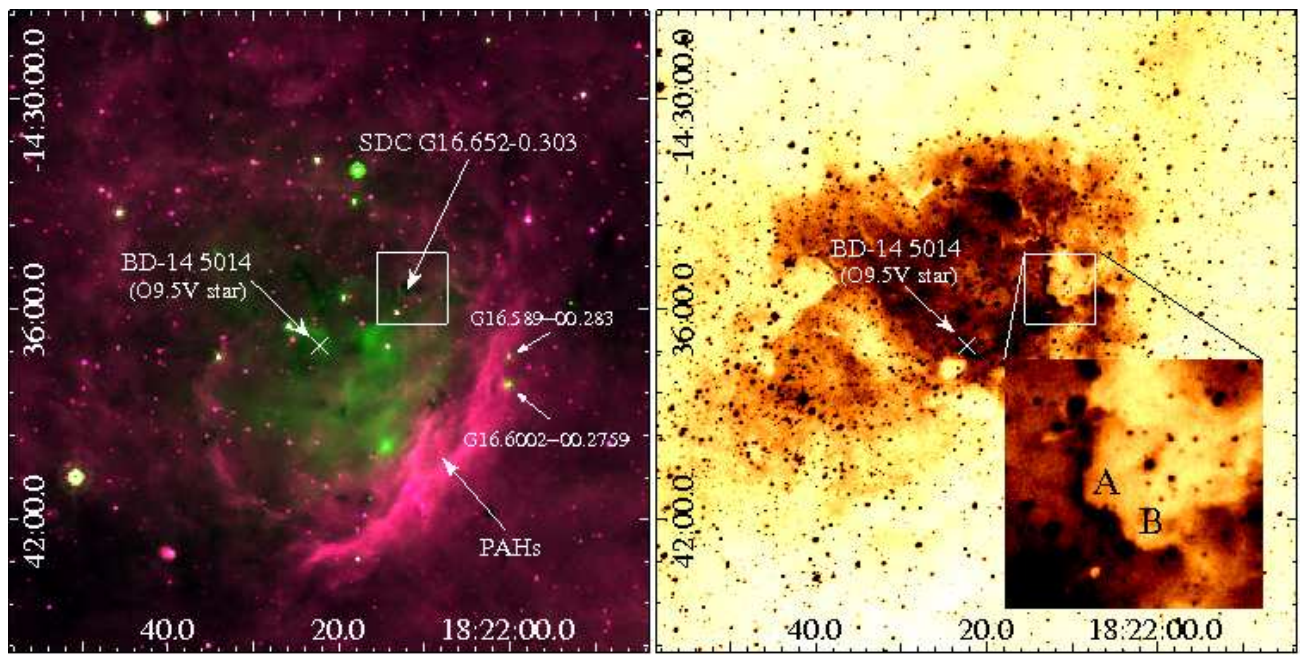}
\caption{Left){\it Spitzer} two-color images (8~$\mu$m in red and
  24~$\mu$m in green) of the infrared dust bubble N18 (infrared
    counterpart of Sh2-48). The white box indicates the region mapped
  with ASTE which includes the BRC. The position of the ionizing star
  BD-14 5014 is shown. Red and green scales go from 60 to 160 and from
  35 to 350~MJy/sr, respectively.  Right) H$_{\alpha}$ image of the
  \ion{H}{ii} region Sh2-48 as obtained from the Super COSMOS H-alpha Survey
  (SHS). A close up view of the BRC is shown where the features A and
  B are indicated. The scale goes from 2600 (white) to 10000~R
  (black).}
\label{description}
\end{figure*}

\section{Introduction}

Bright-rimmed clouds (BRCs) are small dense clouds located at the
border of evolved \ion{H}{ii} regions. The illumination of these dark clumps
by nearby OB stars might be responsible for triggered collapse and
subsequent star formation (e.g. \citealt{sandford1982};
\citealt{ber89}; \citealt{lef94}). The process begins when the
ionization front associated with an \ion{H}{ii} region moves over a
pre-existing molecular condensation, creating a dense outer shell of
ionized gas named: ionized boundary layer (IBL), which surrounds the
rim of the clump.  If the IBL is over-pressured with respect to the
molecular gas within the BRC, shocks are driven into the cloud
compressing the molecular material until the internal pressure is
balanced with the pressure of the IBL (about $10^5$~yr later). At this
stage the collapse of the clump begins a process leading to the
creation of a new generation of stars.  After that, the shock front
dissipates and the cloud is considered to be in a quasi-steady state
known as the cometary globule stage ($\sim 10^6$~yr; \citealt{ber90};
\citealt{lef94}). This mechanism of triggered star formation, known as
radiation-driven implosion (RDI), was first proposed by
\citet{reipurth1983}, and may be responsible for the production of
hundreds of stars in each \ion{H}{ii} region \citep{ogu02}. Finally, the mass
loss resulting from photo-evaporative processes ultimately leads to
the destruction of the cloud on a timescale of several million years
\citep{meg97}.

Star formation in BRCs has long been suspected
(e.g. \citealt{Wootten1983}). \citet{sug91} and \citet{sug94} compiled
catalogs (the so-called SFO Catalog) of 44 BRCs in the northern sky
and 45 BRCs in the southern sky, each associated with an IRAS point
source of low dust temperature. Near-IR imaging observations by
\citet{sugitani1995} indicated that BRCs are often associated with a
small cluster of young stars showing not only an asymmetric spatial
distribution with respect to the cloud but also a possible age
gradient. \citet{sugitani2000} report that young stellar objects
(YSOs) detected inside BRCs tend to lie close to the line joining the
center of mass of the cloud and the ionizing stars.  Detailed
characterization of the physical properties of some BRCs included in
the SFO Catalog were made based on submillimeter and radio continuum
observations (e.g. \citealt{morgan2004}; \citealt{thompson2004};
\citealt{urq06}; \citealt{morgan2008} and references therein). Some
authors concluded that a radiative-driven implosion mechanism is in
progress in many (but not all) of the SFO BRCs and that massive stars
are being formed there \citep{urquhart2009}.

In this work, we present new molecular line data towards a BRC
associated with the \ion{H}{ii} region Sh2-48, obtained using the Atacama
Submillimeter Telescope Experiment (ASTE) and radio continuum
observations at 5~GHz carried out using the Jansky Very Large
  Array (JVLA). We characterize the molecular clump and its
associated IBL, and investigate the balance pressure between the
molecular and the ionized gas using our observations which, when
combined with public infrared (IR) data, provide a comprehensive
picture of the star formation process associated with this BRC and
allow us to discern whether or not it was triggered by the proposed
mechanism.

\begin{figure*}
\centering
\includegraphics[width=13cm]{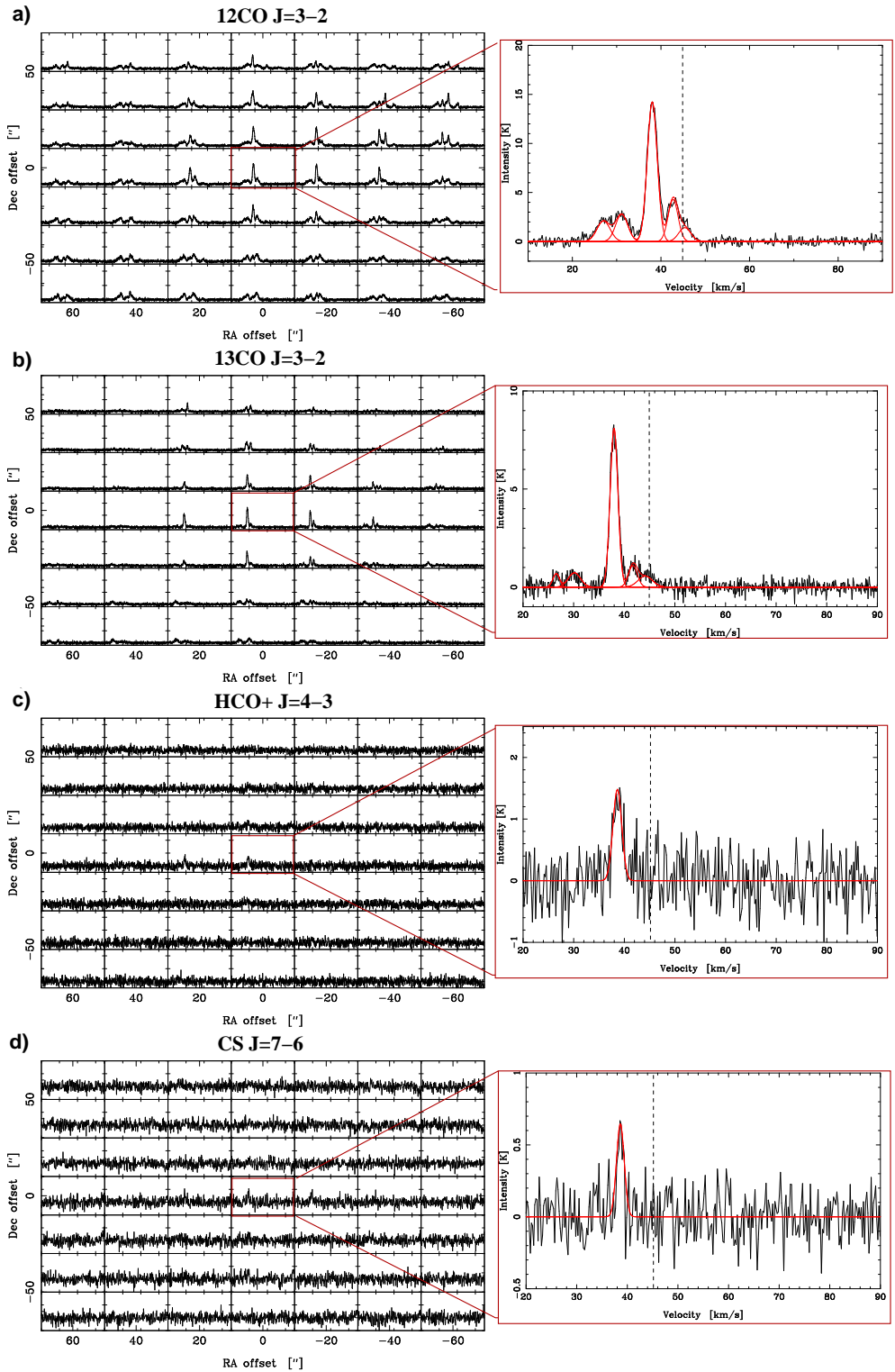}
\caption{Left-column: $^{12}$CO J=3$-$2 (a), $^{13}$CO J=3$-$2
    (b), HCO$^+$ J=4$-$3 (c), and CS J=7$-$6 (d) spectra obtained
    towards the 2$^{\prime}\times 2^{\prime}$ region (white box in
    Fig. \ref{description}) mapped with ASTE. Right-column: spectra
    towards the position (0, 0) of the four transitions smoothed to a
    velocity resolution of 0.22~kms$^{-1}$. The single or
    multiple-component Gaussian fits are shown in red. The dashed line
    marks the systemic velocity of the molecular cloud. All velocities
    are in local standard of rest.}
\label{spectra4lines}
\end{figure*}

\subsection{Presentation of a BRC embedded in the \ion{H}{ii} region Sh2-48}

Sh2-48 is an irregular \ion{H}{ii} region about 10$^{\prime}$~in
size, centered at RA = 18$^{\rm h}$22$^{\rm m}$24.1$^{\rm s}$,
dec.=-14$^{\circ}$35$^{\prime}$09$^{\prime\prime}$(J2000) which was
first catalogued by \citet{sharpless1959}. \citet{avedisova1984}
identified the star BD-14 5014 with spectral type O9.5V
(\citealt{vogt1975}; \citealt{vijapurkar1993}) as the exciting star of
Sh2-48.  \citet{loc89}, based on radio recombination lines, estimated
for Sh2-48 a radial velocity of $\sim$ 44.9~kms$^{-1}$, while
\citet{blitz1982} in their Catalog of CO radial velocities toward
Galactic \ion{H}{ii} regions reported a molecular cloud related to Sh2-48 at
44.6~kms$^{-1}$. Later, \citet{anderson2009} detected molecular gas
associated with the \ion{H}{ii} region at a radial velocity of about
43.2~kms$^{-1}$, which using a flat rotation model for our Galaxy
(with R$_{\odot}=7.6 \pm 0.3$~kpc and $\Theta_{\odot}=214 \pm
7$~kms$^{-1}$) corresponds to the near and far distances of about 3.8
and 12.4~kpc, respectively. Based on \ion{H}{i} absorption studies,
the authors resolved the ambiguity in favor of the far
distance. However, several works based on spectrophotometry studies of
BD-14 5014 (e.g. \citealt{vogt1975}; \citealt{crampton1978};
\citealt{avedisova1984}) established better constraints in favor of
the near distance for the ionizing star of Sh2-48. In what follows we
adopt 3.8~kpc as the most likely distance to Sh2-48 and its associated
BRC.

The Infrared Dust Bubbles Catalog compiled by \citet{churchwell2006},
includes the bubble N18, which can be identified as the \ion{H}{ii}
region Sh2-48 with its associated photodissociation region (PDR; seen
at 8~$\mu$m). In Figure \ref{description}-left we present a composite
two-color image (8~$\mu$m in red and 24~$\mu$m in green) of N18. N18
is an open infrared dust bubble with spiraling filaments at 8~$\mu$m
that partially encircle the emission at 24~$\mu$m, mainly related to
the small dust grains. On the western border of the bubble,
\citet{deharveng2010} identified two compact radio sources with
associated 24~$\mu$m emission, G16.6002--00.2759 and G16.589--00.283,
suggesting that they are likely compact \ion{H}{ii} regions whose
association with N18 is uncertain. In this paper, we focus our
attention in a particularly interesting region (delimited in
Fig. \ref{description} with a white box) where we identify a new
BRC. This BRC has not any IRAS point source associated, and has not
been included in the SFO Catalog. As seen in projection, it appears to
be embedded within Sh2-48 towards the northwestern border of the
\ion{H}{ii} region. The associated bright rim, better appreciatted at
8~$\mu$m, delineates the border of the catalogued {\it Spitzer} dark
cloud SDC G16.652-0.303 which is facing the ionizing star.  The curved
morphology of the illuminated bright rim suggests that the action of
the nearby O star must have shaped the dark molecular cloud.

Figure \ref{description}-right shows an image of the H$_{\alpha}$
emission arising from the ionized gas associated with Sh2-48 as
obtained from the Super COSMOS H-alpha Survey (SHS). The optical image
highlights the illuminated border of the BRC (see white box). The
bright rim is clearly facing the ionizing star and preceding a region
of high visual extinction associated with the above mentioned {\it
  Spitzer} dark cloud SDC G16.652-0.303. The ionized gas appears to be
located onto the surface of the BRC. A close up view of the BRC
included in Fig. \ref{description}-right, shows two curved features,
labeled A and B, which may have been sculpted by the action of the
\ion{H}{ii} region. In particular, the bright rim associated
with the feature A is the brightest one, which is in agreement with
the fact that it is the only one detected at 8~$\mu$m.

\section{Observations and data reduction}

\subsection{Molecular observations}

The molecular line observations were carried out on June 12 and 13,
2011 with the 10m Atacama Submillimeter Telescope Experiment
\citep[ASTE;][]{Ezawa04}. We used the CATS345~GHz band receiver, which
is a two-single band SIS receiver remotely tunable in the LO frequency
range of 324-372~GHz. We simultaneously observed $^{12}$CO J=3$-$2 at
345.796~GHz and HCO$^+$ J=4$-$3 at 356.734~GHz, mapping a region of
2$^{\prime}\times 2^{\prime}$~ centered at RA = 18$^{\rm h}$22$^{\rm
  m}$11.39$^{\rm s}$,
dec.=-14$^\circ$35$^{\prime}$24.81$^{\prime\prime}$ (J2000). We also
observed $^{13}$CO J=3$-$2 at 330.588~GHz and CS J=7$-$6 at
342.883~GHz towards the same region. The mapping grid spacing was
20$^{\prime\prime}$~in both cases and the integration time was 20~sec
($^{12}$CO and HCO$^+$) and 40~sec ($^{13}$CO and CS) per
pointing. All the observations were performed in position switching
mode. We verified that the off position was free of emission.  We used
the XF digital spectrometer with a bandwidth and spectral resolution
set to 128~MHz and 125~kHz, respectively. The velocity resolution was
0.11~kms$^{-1}$ and the half-power beamwidth (HPBW) was about
22$^{\prime\prime}$~for all observed molecular lines. The system
temperature varied from T$_{\rm sys}$ = 150 to 200~K. The main beam
efficiency was $\eta_{\rm mb} \sim$0.65. All the spectra were Hanning
smoothed to improve the signal-to-noise ratio. The baseline fitting
was carried out using second order polynomials for the $^{12}$CO and
$^{13}$CO transitions and third order polynomials for the HCO$^+$ and
CS transitions. The polynomia were the same for all spectra of the map
at a given transition.  The resulting rms noise of the observations
was about 0.2~K for $^{13}$CO J=3$-$2 and CS J=7$-$6, and about 0.4~K
for $^{12}$CO J=3$-$2 and HCO$^+$ J=4$-$3 transitions.

\subsection{Radio continuum observations}

The radio continuum observations towards the bright-rimmed cloud were
performed in a single pointing with the Karl G. Jansky Very Large
Array (JVLA) in its C configuration, on 2012 February 7 and 12
(project ID:12A-020) for a total of 70 minutes on-source integration
time. We used the wideband 4-8~GHz receiver system centered at 4.7 and
7.4~GHz, which consists in 16 spectral windows with a bandwidth of
128~MHz each, spread into 64 channels. Data processing was carried out
using the CASA and Miriad software packages, following standard
procedures.  The source J1331+3030 was used for primary flux density
and bandpass calibration, while phases were calibrated with
J1820-2528. For the scope of this paper we only reconstructed an image
centered at 5~GHz with a bandwith of 768~MHz using the task MAXEN in
MIRIAD, which performs a maximum entropy deconvolution algorithm on a
cube. The resulting synthesised beam has a size of
7$^{\prime\prime}.5 \times 5^{\prime\prime}$.4, and the rms noise of
the final map is 0.016~mJy/beam.

The observations are complemented with H$_{\alpha}$, near- and mid-IR
data extracted from public databases and catalogs, which are described
in the corresponding sections.

\section{Results and Analysis}

\subsection{The molecular gas}
\label{molecular}

\begin{figure}
\centering
\includegraphics[width=7.5cm]{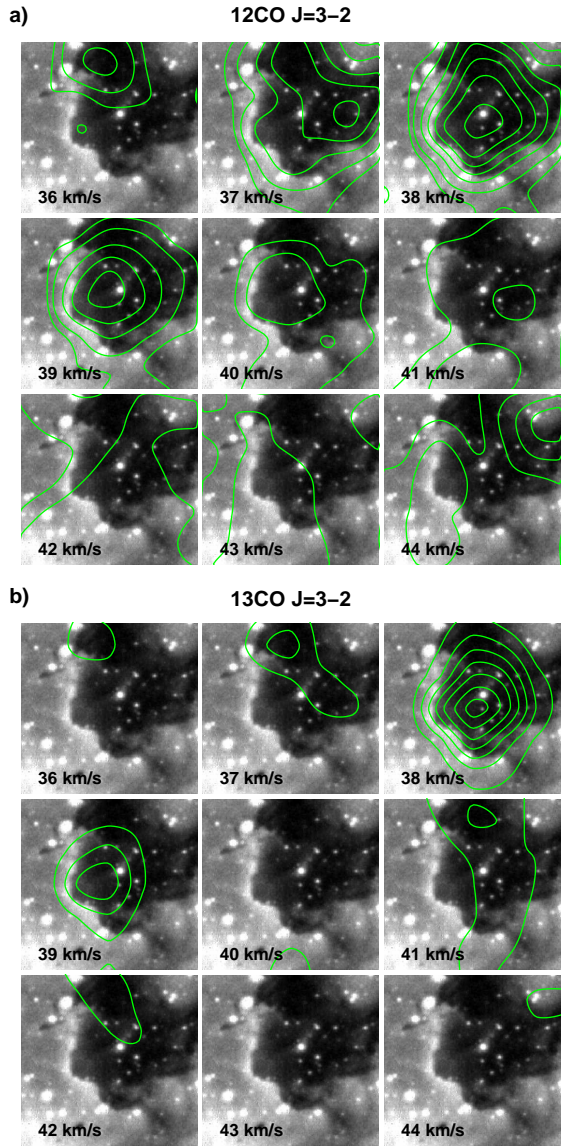}
\caption{Velocity channel maps of the $^{12}$CO J=3$-$2 (a) and
    $^{13}$CO J=3$-$2 (b) emission from 35 to 44~kms$^{-1}$ integrated
    every 1~kms$^{-1}$ (green contours) superimposed onto the
    H$_{\alpha}$ emission of the BRC. The given velocities correspond
    to the higher velocity of each interval. Contours are plotted
    above the 5$\sigma$ of the rms noise level.}
\label{velmapco}
\end{figure}

 Figure \ref{spectra4lines}(left-column) shows the $^{12}$CO
  J=3$-$2 (a), $^{13}$CO J=3$-$2 (b), HCO$^+$ J=4$-$3 (c), and CS
  J=7$-$6 (d) spectra obtained towards the
  2$^{\prime}\times2^{\prime}$ analyzed region (white box in
  Fig. \ref{description}). The corresponding profiles towards the
  positions (0, 0) are also shown in the right-column of the figure.
  The $^{12}$CO J=3$-$2 profile towards the (0, 0) offset exhibits a
  quintuple peak structure with components centered at about 27, 31,
  38, 42, and 45~kms$^{-1}$. In particular, the velocity component
  centered at about 38~kms$^{-1}$, which is detected towards the
  central region, is most intense at this position, while the
  component centered at about 45~kms$^{-1}$, whose velocity coincides
  with the systemic velocity of the molecular cloud related to Sh2-48
  is most intense at the position ($-$40, $+$40). The $^{13}$CO
  J=3$-$2 spectrum at the (0, 0) position exhibits a behavior similar
  to that of $^{12}$CO J=3$-$2 with the same five velocity components
  centered at 27, 31, 38, 42, and 45~kms$^{-1}$. As in the case of the
  $^{12}$CO J=3$-$2 emission, the component centered at 38~kms$^{-1}$~
  is most intense at this position and the intensity maximum of the
  component at 45~kms$^{-1}$ is at the ($-$40, $+$40) offset. The
  HCO$^+$ J=4$-$3 line has a single velocity component above 3$\sigma$
  centered at 38.4~kms$^{-1}$ towards the (0, 0) position. Finally,
  the CS J=7$-$6 spectrum at the (0, 0) offset also shows a single
  velocity component centered at about 38.5~kms$^{-1}$ above 3$\sigma$
  of the rms noise level. The detection of this molecular transition
  reveals the presence of warm and dense gas mapping the dense core of
  the cloud. The velocity component related to the BRC is 6~kms$^{-1}$
  blue-shifted respect to the systemic velocity of the parental
  molecular cloud in which it is embedded. This fact suggests that the
  molecular clump has probably been pushed forward by the O star and
  currently it is moving in our direction with respect to the centre
  of the complex. This result is in agreement with the predictions of
  the works of \citet{pittard2009} and \citet{mizuta2006}, which based
  on simulations of radiative and shock destruction of clouds, have
  shown that the head of a pillar in a molecular cloud exposed to the
  action of a neighboring massive star, is accelerated outward or
  evaporated. As such, the position of the pillar head is offset from
  the initial cloud position.

In Figure \ref{velmapco} we show the velocity channel maps of the
$^{12}$CO J=3$-$2 (a) and the $^{13}$CO J=3$-$2 (b) emission
distributions from 35 to 44~kms$^{-1}$, integrated every 1~kms$^{-1}$
(green contours) superimposed onto the H$_{\alpha}$ image of the
BRC. As can be seen from this figure, the $^{12}$CO J=3$-$2 emission
distribution is well correlated with the BRC between 36 and
40~kms$^{-1}$ and the $^{13}$CO J=3$-$2 emission around 38~kms$^{-1}$
exhibits an excellent morphological correlation with the BRC as
seen in the optical image.

\begin{figure}
\centering
\includegraphics[width=7.5cm]{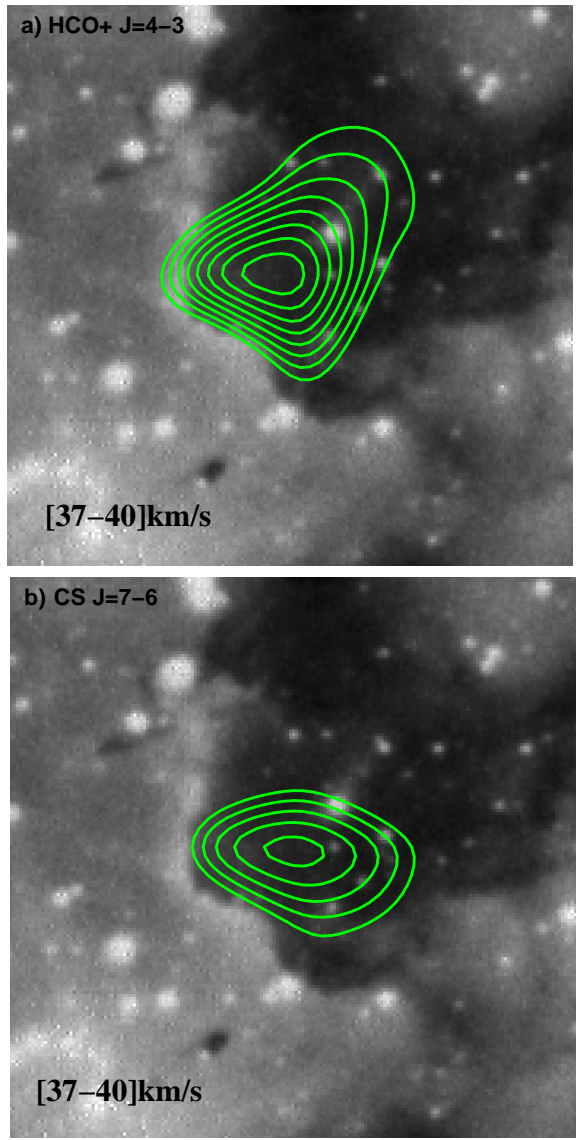}
\caption{a) Emission distribution of the HCO$^+$ J=4$-$3
    transition integrated between 37 and 40~kms$^{-1}$ (green
    contours) superimposed onto the H$_{\alpha}$ emission of the BRC.
    Contours levels go from 0.8 to 2.2~K~kms$^{-1}$ in steps of 0.2 ~K
    kms$^{-1}$. b) Emission distribution of the CS J=7$-$6 transition
    integrated between 37 and 40~kms$^{-1}$ (green contours)
    superimposed onto the H$_{\alpha}$ emission of the BRC.  Contours
    levels are at 0.3, 0.4, 0.5, 0.6, and 0.7~K~kms$^{-1}$.}
\label{hco-cs_mapvel}
\end{figure}

Figure \ref{hco-cs_mapvel} shows the HCO$^+$ J=4$-$3 (a) and the CS
J=7$-$6 (b) emission distributions above 3$\sigma$ of the rms noise
level integrated between 37 and 40~km$s^{-1}$ (green contours)
superimposed onto the H$_{\alpha}$ image of the BRC. From this figure
it can be seen the good morphological correlation between the HCO$^+$
J=4$-$3 emission distribution with the curved H$_{\alpha}$ features of
the BRC. Besides, the intensity gradient seems to be steeper in the
direction of the ionizing star, suggesting a possible compression on
the molecular gas. The molecular clump detected at the CS J=7$-$6
transition is slightly set back from the bright rim with respect to
the direction of the ionizing star. The maximum of the emission of
both lines, HCO$^+$ J=4$-$3 and CS J=7$-$6, are positionally
coincident.

Table \ref{tab_fit} lists the emission peaks parameters derived from a
Gaussian fitting for the four molecular transitions on the position
(0, 0). V$_{LSR}$ represents the central velocity referred to the
Local Standard of Rest, T$_{mb}$ the peak brightness temperature, and
$\Delta$v the line FWHM. Errors are formal 1$\sigma$ value for the
model of Gaussian line shape. All velocities are in the local
  standard of rest.

\begin{table} 
\caption{Emission peaks parameters derived from a Gaussian fitting for
  the four molecular transitions on the position (0, 0).} \centering
\begin{tabular}{cccc}
\hline 
Transition & V$_{LSR}$ [kms$^{-1}$] &  T$_{mb}$ [K] & $\Delta$v [kms$^{-1}$]\\
\hline 
CS J=7--6 & 38.5$\pm$0.9 & 0.6$\pm$0.2 & 1.8$\pm$0.6\\
         
\hline
HCO$^+$ J=4--3  & 38.4$\pm$0.4 & 1.5$\pm$0.2 & 2.0$\pm$0.5\\
               
\hline 

$^{13}$CO J=3--2 & 26.9 $\pm$ 0.6 & 0.7 $\pm$ 0.2 & 1.5 $\pm$ 0.6\\ 
                       & 30.7 $\pm$ 0.4 & 0.8 $\pm$ 0.3 & 1.8 $\pm$ 0.6\\ 
                       & 38.1 $\pm$ 0.4 & 8.1 $\pm$ 0.3 & 1.9 $\pm$ 0.7\\ 
                       & 41.8 $\pm$ 0.6 & 1.3 $\pm$ 0.4 & 1.5 $\pm$ 0.4\\ 
                       & 44.6 $\pm$ 0.5 & 0.6 $\pm$ 0.4 & 3.1 $\pm$ 0.4\\  
\hline 
$^{12}$CO J=3--2 & 26.9$\pm$0.1 & 2.2$\pm$1.1 & 3.6$\pm$0.4\\ 
                       & 30.8$\pm$0.2 & 2.9$\pm$0.8 & 3.6$\pm$0.6\\ 
                       & 37.9$\pm$0.6 & 14.1$\pm$0.7 & 2.9$\pm$0.5\\ 
                       & 42.2$\pm$0.4 & 4.2$\pm$0.2 & 2.8$\pm$0.4\\ 
                       & 45.1$\pm$0.5 & 1.2$\pm$0.2 & 3.3$\pm$0.4\\
\hline
\label{tab_fit}
\end{tabular}
\end{table}

\subsection{Column density and mass estimates of the molecular clump}
\label{param_clump}

Assuming LTE conditions we estimate the $^{13}$CO J=3--2 opacity,
$\tau_{13}$, based on the following equation:

\begin{equation}
\small 
\tau_{13}=-ln\left(1-\frac{T_{peak}(^{13}{\rm CO})}{T_{peak}(^{12}{\rm CO})}\right)
\end{equation}

\noindent where the $T_{peak}$ at 38~kms$^{-1}$ was measured at the position (0, 0) for
both transitions. We obtain $\tau_{13} \sim$ 0.8, which suggests that
the $^{13}$CO J=3$-$2 line can be considered optically thin in this
molecular condensation.

The excitation temperature, $T_{ex}$, of the $^{13}$CO J=3$-$2 line is
estimated from:

\begin{equation}
T_{peak}(^{13}{\rm CO})=\frac{h\nu}{k} \left(\frac{1}{e^{h\nu/k
    T_{ex}}-1}-\frac{1}{e^{h\nu/k T_{BG}}-1}\right) \times
(1-e^{-\tau_{13}})
\end{equation}

\noindent where for this transition $h\nu/k=15.87$. Assuming $T_{BG}$
= 2.7~K, and considering the $T_{peak}$($^{13}$CO) = 8~K, we derive a
$T_{ex} \sim$ 21~K for this line. This value is significantly higher
than would be expected for a starless clump ($T \leq 10$~K;
\citealt{shinnaga2004}), evidencing that the molecular clump has an
internal heating mechanism. Then, we derive the $^{13}$CO column
density from (see e.g. \citealt{buc10}):

\begin{equation}
{\rm N}(^{13}{\rm CO})=8.28 \times 10^{13}
e^{\frac{15.87}{T_{ex}}}\frac{T_{ex}+0.88}{1-exp(\frac{-15.87}{T_{ex}})}\int{\tau_{13} {\rm dv}}
\end{equation}

\noindent where, taking into account that $^{13}$CO J=3$-$2 transition
can be considered optically thin, we use the approximation:

\begin{equation}
\int{\tau {\rm dv}}=\frac{1}{J(T_{ex})-J(T_{BG})}\int{\rm{T_{mb}} {\rm dv}}
\end{equation}

\noindent with

\begin{equation}
J(T) = \frac{h\nu/k}{exp(\frac{h\nu}{kT})-1}.
\end{equation}

\noindent From the estimated N($^{13}$CO)$\sim 8 \times
  10^{15}~{\rm cm}^{-2}$, and assuming the [H$_2$]/[$^{13}$CO] =
  77$\times 10^4$ ratio \citep{wil94}, we derive an H$_2$ column
density, N(H$_2$)$\sim 6 \times 10^{21}~{\rm cm}^{-2}$. Using the
relation $M=\mu m_H d^2 \Omega {\rm N(H_2)}$, where $\mu$ is the mean
molecular weight per H$_2$ molecule ($\mu \sim 2.72$), $m_H$ the
hydrogen atomic mass, $d$ the distance, and $\Omega$ the solid angle
subtended by the structure, then the total mass of the clump turns out
to be $M \sim 180$~M$_\odot$ and the volume density, n(H$_2$)$\sim 3
\times 10^3 {{\rm cm}^{-3}}$. The errors in these estimates are about
30\% and 40\%, respectively.

We also independently calculate the mass and the volume density of the
clump based on the associated dust continuum emission. In particular,
we use the integrated flux of the continuum emission at 1.1~mm as
obtained from The Bolocam Galactic Plane Survey II Catalog (BGPS II;
\citealt{ros10}). Following \citet{be02} and \citet{hild83}, and
considering the flux estimated with the aperture of 80$^{\prime\prime}$, to be
comparable in size with the BRC, we calculate the mass of the clump in solar masses
from:

\begin{eqnarray}
M_{gas}=\frac{1.3 \times 10^{-3}}{J_{\nu}(T_{dust})}\frac{a}{0.1
  \mu{\rm m}}\frac{\rho}{3 {\rm g
    cm}^{-3}}\frac{R}{100}\frac{F_{\nu}}{{\rm
    Jy}}\nonumber\left(\frac{d}{{\rm kpc}}\right)^2
\left(\frac{\nu}{2.4 {\rm THz}}\right)^{-3-\beta} 
\end{eqnarray}

\noindent where $J_{\nu}(T_{dust}$) = [exp($h\nu/kT_{dust})-1]^{-1}$
and $a, \rho, R,$ and $\beta$ are the grain size, grain mass density,
gas-to-dust ratio, and grain emissivity index for which we adopt the
values of 0.1~$\mu$m, 3~g cm$^{-3}$, 100, and 2, respectively
(\citealt{hun97}, \citealt{hun00}, and \citealt{mol00}). Assuming a
dust temperature of 20~K and considering the integrated flux intensity
$S_{80}$ = 0.285~Jy at 1~mm \citep{ros10}, we obtain $M_{gas} \sim
260$~M$_\odot$ and a volume density, $n$(H$_2) \sim 3 \times
10^3$~cm$^{-3}$, in good agreement with those values derived from the LTE
calculations using the $^{12}$CO J=3$-$2 and $^{13}$CO J=3$-$2
transitions.

\subsection{The ionized boundary layer associated with the BRC}
\label{IBL}

\begin{figure}
\centering \includegraphics[width=8cm]{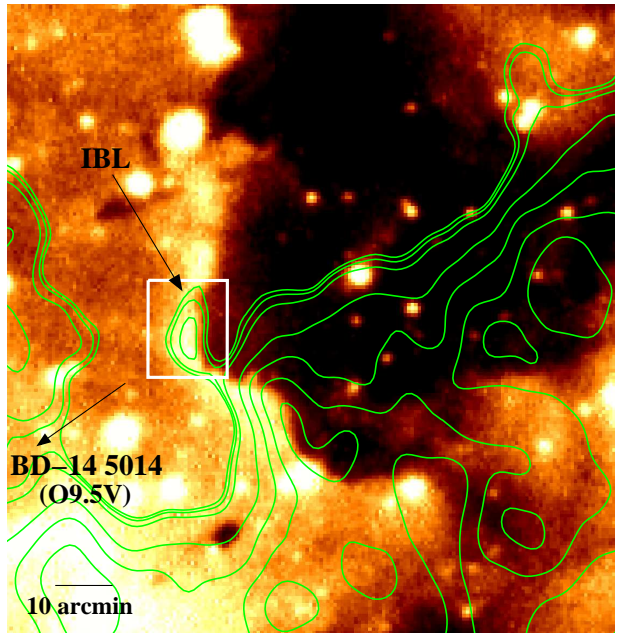}
\caption{H$_{\alpha}$ image of the BRC. The green contours represent
  the radio continuum emission at 5~GHz. Contours levels are at 0.7
  (about 4$\sigma$ above the rms noise level), 0.9, 1, 2, 3, 4, and 5
  $\times 10^{-4}$~Jy/beam. The white box indicates the IBL region
  associated with the protrusion A.}
\label{BR_radio}
\end{figure}

Figure \ref{BR_radio} shows the new JVLA radio continuum emission at
5~GHz (green contours) superimposed onto the H$_{\alpha}$ image of the
BRC. At first glance, it can be appreciate the radio continuum
emission related to Sh2-48, extending over most of the studied
region. A noticeable radio feature is the arc-like radio filament that
perfectly matches the optical emission of the bright rim related to
the curved feature A. This positional and morphological correlation
suggests that this radio continuum emission arises from the ionized
gas located in the illuminated border of the molecular clump. Thus,
the radio continuum emission allow us to estimate the ionizing photon
flux impinging upon the illuminated face of the BRC and the electron
density of the ionized boundary layer. We estimate the radio continuum
flux density of the arc-like radio filament at 5~GHz in $2.3 \times
10^{-4}$ Jy.

Assuming that all of the ionizing photon flux is absorbed within the
IBL, we determine the photon flux, $\Phi$, and the electron density,
$n_e$, using the equations detailed by \citet{lefloch1997} and
modified by \citet{thompson2004}:

\begin{equation}
\Phi = 1.24 \times 10^{10} S_{\nu} T_e^{0.35}\nu^{0.1} \theta^{-2} [{\rm cm}^{-2}{\rm s}^{-1}]
\end{equation}

\noindent

\begin{equation}
n_e=122.21 \times \left(\frac{S_{\nu}T_e^{0.35}\nu^{0.1}}{\eta R
  \theta^2}\right)^{1/2} [{\rm cm}^{-3}]
\end{equation}

\noindent where S$_{\nu}$ is the integrated flux density in mJy, $\nu$
is the frequency in GHz, $\theta$ is the angular diameter over which
the flux density is integrated in arc-seconds, $\eta R$ is the shell
thickness in pc, and $T_e$ is the electron temperature in K.  Assuming
an average electron temperature of about $10^4$~K and an $\eta$ = 0.2
\citep{ber89}, and considering an effective $\theta$ of about 12~$^{\prime\prime}$ and
a clump radius $R \sim 0.6^{\prime}$~(about 0.67 pc at the distance of
3.8~kpc), we derive the photon flux and electron density values of
$\Phi \sim 5.8 \times 10^8 {\rm cm}^{-2}{\rm s}^{-1}$ and $n_e \sim
73~{\rm cm}^{-3}$, respectively. The main sources of error in the
electron density estimate come from the assumption on $\eta$ and from
the uncertainty in the distance, which combined give an error of about
40\%.

The mean electron density value obtained for the IBL is almost a
factor three greater than the critical value of $n_e \sim 25 {\rm
  cm}^{-3}$, above which an IBL is able to develop around a molecular
clump \citep{lef94}. This fact reinforces the hypothesis that the
clump is being photoionized by the nearby O9.5V star. However, it is
not clear to what extent the ionization has influenced the evolution
of the cloud, and what role, if any, has played in triggering star
formation.

Finally, we also calculate the predicted ionized flux, $\Phi_{pred}$,
of the IBL considering a Lyman photon flux of about $1.8 \times
10^{48}$~ph s$^{-1}$ \citep{sch97} for the ionizing star BD-14 5014
which is located at least 3~pc away from the clump. The predicted
ionized photon flux, $\Phi_{pred} \sim 16.7 \times 10^8 {\rm
  cm}^{-2}{\rm s}^{-1}$ is three times greater than the value
estimated from the radio continuum observations, which is not
surprising given that the former is a strict upper limit due to
projection effects and dust absorption.

\section{Discussion}

\subsection{Testing the RDI mechanism through a pressure balance analysis}
\label{pressure}

To evaluate the pressure balance between the ionized gas of the IBL
and the neutral gas of the molecular cloud we use the results of
Sections \ref{param_clump} and \ref{IBL}. The analysis of the balance
between the external and internal pressures, $P_{ext}$ and
$P_{int}$, respectively, gives a good piece of information about the
influence that the ionization front has had in the evolution of the
BRC.  Following \citet{thompson2004} the pressures are defined as:

\begin{equation}
P_{int} \simeq \sigma^2 \rho_{int}
\end{equation}

\begin{equation}
P_{ext} = 2 \rho_{ext} c^2
\end{equation}

\noindent where $\sigma^2$ is the square of the velocity dispersion,
defined as $\sigma^2$= $\Delta v^2 /(8~{\rm ln}2)$, with $\Delta v$
the line width of the $^{13}$CO J=3$-$2 line taken from the profile at
the (0, 0) offset, $\rho_{int}$ is the clump density, $\rho_{ext}$ the
ionized gas density, and $c\sim$11.4~kms$^{-1}$ (e.g. \citealt{urq06}) a
typical sound speed for these regions. To estimate the ionized gas
pressure for the IBL we use the electron density calculated in
Sect. \ref{IBL}, obtaining $P_{ext}/k_B \sim 23 \times
10^5$~cm$^{-3}$~K, which is among the lowest values estimated in
similar regions (see e.g. \citealt{morgan2004}).  To estimate the
internal pressure of the molecular clump, we use the H$_2$ volume
density, n(H$_2$), calculated in Sect. \ref{param_clump}, yielding a
$P_{int}/k_B \sim 5 \times 10^5$~cm$^{-3}$~K.

The comparison between both pressures reveals that the clump is under
pressure by a factor four with respect to its IBL, suggesting that the
shocks are currently being driven into the surface layers. In this
way, the BRC could be in a pre-pressure balance state,  where the
  \ion{H}{ii} region has only recently begun to affect the molecular gas and
  shocks have not propagated very far into the clump. Thus, it is
  likely that the molecular clump predate the arrival of the
  ionization front.

\subsection{Young stellar object population}
\label{yso_population}

\begin{figure}
\centering \includegraphics[width=8cm]{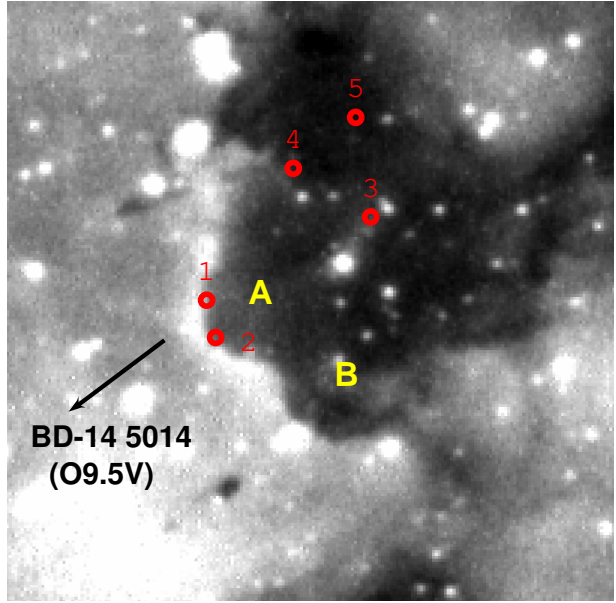}
\caption{Spatial distribution of the YSO candidates towards the BRC. Their positions are indicated with the red
  circles. The black arrow shows the direction of the ionizing star.}
\label{ysos}
\end{figure}

Given the result of previous section, any young stellar object (YSO)
triggered via the radiation-driven implosion mechanism should be
placed at the illuminated border of the BRC. In this section we look
for YSO candidates associated with the molecular condensation. YSOs
use to be classified based on their evolutive stage: class I are the
youngest sources which are still embedded in dense envelopes of gas
and dust, and class II are those sources whose emission is mainly
originated in the accretion disk surrounding the central protostar.
In both cases, a YSO will exhibit an infrared excess which is mainly
due to the presence of the envelope and/or the disk of dust around the
central object, but not attributed to the scattering and absorption of
the interstellar medium along the line of sight. In other words, YSOs
are intrinsically red sources.  \citet{rob08} defined an infrared
color criterion to identify intrinsically red sources based on {\it
  Spitzer} data. They must satisfy the condition $[4.5]-[8.0] \geq 1$,
where [4.5] and [8.0] are the magnitudes in the 4.5 and 8.0~$\mu$m
bands, respectively.  Following this criterion we find five
intrinsically red sources towards the region of the BRC and its
surroundings. In Table \ref{ysos_table} we report the magnitudes of
the intrinsically red sources in the {\it Spitzer}-IRAC bands
(Col. 3-6), specifying the GLIMPSE designation (Col. 2) and the
\citet{all04} classification (Col. 7), and in Figure \ref{ysos} we
show their location. Sources \#1 and \#2 are located at the border of
the bright rim labeled A, they are spatially separated from the other
three embedded sources, and are young class I YSOs. All these
characteristics support the possibility that the RDI processes have
triggered their formation. However, we can not discard that both YSOs
have been formed previously by other mechanism and now are being
unveiled by the advancing ionization front.  Given that the BRC
  would be located in the near part of the molecular complex, due to
  projection effects we can not rule out that the YSOs \#1 and \#2 are
  currently detached from the clump. In any case, if the newly born
  stars are massive, their feedback may ultimately destroy their natal
  molecular cloud shutting off star formation or may simultaneously
  trigger the birth of new generations of stars, as in the scenario of
  progressive star formation in the Carina nebula described by
  \citet{smith2010}. Depending on how massive the recently formed
  stars are, they will affect more or less the local environment of
  the BRC. Future works tending to characterize these sources will be
  relevant to disentangle such effects.

Concerning the YSOs \#3, \#4, and \#5, the fact that the shocks
induced by the \ion{H}{ii} region Sh2-48 are being driven into
the external layer of the BRC (see Section \ref{pressure}), it is
unlikely that these sources have been triggered via the
radiation-driven implosion mechanism.

\begin{table*} 
\caption{Mid-IR magnitudes and \citet{all04} classification of the
  point sources satisfying the condition $[4.5]-[8.0] \geq 1$
  towards the BRC.}  
\centering
\begin{tabular}{ccccccc}
\hline 
\hline 
Source & GLIMPSE Desig. & 3.6~$\mu$m & 4.5~$\mu$m & 5.8~$\mu$m & 8.0~$\mu$m & Class\\ 

& & (mag) & (mag) & (mag) & (mag) & \\ 
\hline
YSO 1 & G016.6534-00.3062 & 13.140 & 12.744 & 10.871 & 9.663 & I\\

YSO 2 & G016.6511-00.3068 & 11.603 & 11.368 & 10.880 & 10.366 & I\\ 

YSO 3 & G016.6532-00.2948 & 12.798 & 12.911 & 12.187 & 11.314 & I\\ 

YSO 4 & G016.6582-00.2976 & 11.714 & 11.569 & 10.882 & 9.961 & II\\ 

YSO 5 & G016.6592-00.2926 & 14.317 & 13.850 & - & 9.017 & -\\
\hline
\label{ysos_table}
\end{tabular}
\end{table*}

\section{Summary}

We present new molecular observations in the $^{12}$CO J=3$-$2,
$^{13}$CO J=3$-$2, HCO$^+$ J=4$-$3, and CS J=7$-$6 lines using the
Atacama Submillimeter Telescope Experiment (ASTE), and radio continuum
observations at 5~GHz using the Karl Jansky VLA instrument, towards a
new bright-rimmed cloud (BRC) located near the border of the \ion{H}{ii} region Sh2-48.

The molecular observations in the different lines reveal the presence
of a relatively dense clump in very good spatial correspondence with the BRC as
observed in the H$_{\alpha}$ emission.

The high angular resolution and sensitivity radio continuum data have
revealed the presence of an arc-like radio filament in excellent
correspondence with the brightest border of the optical emission of
the BRC, highly suggestive to be the associated ionized boundary layer
(IBL). We derive an electron density for the IBL of about
73~cm$^{-3}$. This value is three times higher than the critical
density above which an IBL can form and be maintained. The location
and morphology of the radio filament, together with the estimate of
the electron density support the hypothesis that the BRC is being
photoionized by the exciting star of Sh2-48.  From the CO and radio
continuum data we estimate the pressure balance between the IBL and
the molecular gas, finding that the BRC is likely to be in a
pre-pressure state.

We have also studied the star formation activity in the region. We
find five YSO candidates embedded in the BRC. Two of them are located
in projection towards the illuminated border of the BRC and are most
probably formed via the RDI mechanism.

\section*{Acknowledgments}

We wish to thank the referee, Dr. Gahm, whose constructive criticism
has helped make this a better paper. M.O., S.P., E.G., and G.D. are
members of the {\sl Carrera del Investigador Cient\'\i fico} of
CONICET, Argentina. This work was partially supported by Argentina
grants awarded by Universidad de Buenos Aires (UBACyT 01/W011),
CONICET and ANPCYT.  M.R. wishes to acknowledge support from FONDECYT
(CHILE) grant No108033. She is supported by the Chilean Center for
Astrophysics FONDAP No. 15010003. The ASTE project is driven by
Nobeyama Radio Observatory (NRO), a branch of National Astronomical
Observatory of Japan (NAOJ), in collaboration with University of
Chile, and Japanese institutes including University of Tokyo, Nagoya
University, Osaka Prefecture University, Ibaraki University, Hokkaido
University and Joetsu University of Education.


\end{document}